\documentclass[conference]{IEEEtran}
\IEEEoverridecommandlockouts
\usepackage{comment}
\usepackage{qcircuit}
\usepackage{xypic}
\usepackage{url}
\usepackage{cite}
\usepackage{amsmath,amssymb,amsfonts}
\usepackage{algorithmic}
\usepackage{graphicx}
\usepackage{textcomp}
\usepackage{xcolor}
\usepackage{booktabs}
\def\BibTeX{{\rm B\kern-.05em{\sc i\kern-.025em b}\kern-.08em
    T\kern-.1667em\lower.7ex\hbox{E}\kern-.125emX}}
\begin{document}

\title{Q-Score: A Quantum-Native Scoring Function for Molecular Docking
%\thanks{Identify applicable funding agency here. If none, delete this.}
}

\author{
\IEEEauthorblockN{Kangyu Zheng}
\IEEEauthorblockA{\textit{Department of Computer Science and Engineering} \\
\textit{The Chinese University of Hong Kong}\\
Shatin, N.T., Hong Kong SAR  \\
kyzheng@cse.cuhk.edu.hk}
\and
\IEEEauthorblockN{Yidong Zhou}
\IEEEauthorblockA{\textit{Department of Electrical and Computer Engineering} \\
\textit{Rutgers University}\\
Piscataway, NJ \\
yidong.zhou@rutgers.edu}
\and
\IEEEauthorblockN{Rui-Hao Li}
\IEEEauthorblockA{\textit{Computational Life Sciences} \\
\textit{Cleveland Clinic}\\
Cleveland, OH \\
LIR9@ccf.org}
\and
\IEEEauthorblockN{Zixin Ding}
\IEEEauthorblockA{\textit{Department of Computer Science} \\
\textit{The University of Chicago}\\
Chicago, IL \\
zixin@uchicago.edu}
\and
\IEEEauthorblockN{Zhiding Liang}
\IEEEauthorblockA{\textit{Department of Computer Science and Engineering} \\
\textit{The Chinese University of Hong Kong}\\
Shatin, N.T., Hong Kong SAR \\
zliang@cse.cuhk.edu.hk}
\and
\IEEEauthorblockN{Shaohua Li}
\IEEEauthorblockA{\textit{Department of Computer Science and Engineering} \\
\textit{The Chinese University of Hong Kong}\\
Shatin, N.T., Hong Kong SAR \\
shaohuali@cuhk.edu.hk}
}

\maketitle

\begin{abstract}
Molecular docking predicts how a small molecule binds to a protein and is a key bottleneck in drug discovery. Classical scoring functions sum empirical pairwise contacts, blind to quantum-mechanical effects like orbital charge transfer that govern binding specificity. We introduce Q-Score, encoding GNN-predicted orbital donor-acceptor energies into a weighted graph and scoring binding by solving a maximum-weight vertex clique problem via Digitized-Counterdiabatic QAOA. Each interaction anchor maps to one qubit and compatibility constraints become edges. Across 11 protein targets, DC-QAOA recovers the exact optimum on 8 at 10 qubits. On 1000 AI-generated molecules, Q-Score is orthogonal to classical scoring with Spearman rho of 0.05, driven by orbital quality with rho of 0.90, and free of molecular-weight bias, enriching for strong orbital interactions at twice the random rate. DC-QAOA achieves a mean approximation ratio of 0.94 with 52 percent exact. Execution of 1000 circuits on IBM Eagle confirms 6-qubit solvability on NISQ hardware.
\end{abstract}


\section{Introduction}
\label{sec:intro}

Molecular docking is one of the most computationally intensive steps in drug discovery, requiring the evaluation of thousands to millions of candidate molecules against a protein target. The task has two components: a search algorithm that proposes candidate placements of a molecule inside the protein binding pocket, and a scoring function that ranks those placements by estimated binding quality. Because scoring is invoked at every step of the search, its cost dominates the overall workflow, and all widely deployed docking tools~\cite{https://doi.org/10.1002/jcc.21334, doi:10.1021/jm0306430} use fast empirical scoring functions that sum pairwise atomic interaction terms such as van der Waals contacts, hydrogen bonds, and desolvation penalties~\cite{doi:10.1021/cr00081a009}.

This additive formulation enables high throughput but introduces a systematic limitation: because the score is a sum over atom pairs, larger molecules accumulate more favorable terms simply by having more atoms, regardless of the quality of their individual interactions~\cite{doi:10.1021/ci600471m}. The scoring function is also unable to capture quantum-mechanical effects, such as orbital-level charge transfer between donor and acceptor groups, $\pi$-stacking stabilization, and hyperconjugation, that often govern binding specificity~\cite{RAHA2007725}. These stereoelectronic effects determine why a molecule binds to one pocket and not another, yet they have no direct representation in the empirical scoring vocabulary.

Recent work has shown that molecular docking can be reformulated as a combinatorial optimization problem by representing candidate protein-ligand interaction contacts as nodes in a weighted graph and encoding geometric compatibility as edges~\cite{PhysRevApplied.21.034036, doi:10.1021/acs.jctc.4c00067}. Selecting the best subset of mutually compatible contacts then becomes a maximum-weight vertex clique problem (MWVCP), which maps naturally onto an Ising Hamiltonian and can be solved using the Quantum Approximate Optimization Algorithm (QAOA)~\cite{farhi2014quantumapproximateoptimizationalgorithm}. However, prior formulations rely on manual pharmacophore annotations and coarse-grained interaction categories to assign node weights, limiting the chemical information encoded in the graph.

In this work, we introduce \emph{Q-Score}, a scoring function that replaces empirical pairwise sums with quantum-chemistry-informed orbital interaction energies solved via quantum optimization. Q-Score uses a graph neural network~\cite{Boiko2025} to predict second-order perturbation energies $E^{(2)}$ from Natural Bond Orbital theory~\cite{B1RP90011K}, encodes the strongest donor-acceptor interactions into an Orbital Interaction Graph (OIG), and obtains the score by solving the resulting MWVCP with Digitized-Counterdiabatic QAOA (DC-QAOA)~\cite{PhysRevResearch.4.013141}. Because each interaction anchor maps to one qubit and geometric compatibility constraints become graph edges, the problem has a natural and compact quantum representation.

We validate Q-Score in two experimental settings. First, in a \emph{redocking} experiment across 11 protein-ligand co-crystal structures, we verify that DC-QAOA recovers the exact optimal orbital interaction set on 8 out of 11 targets at 10 qubits, with the selected cliques correctly identifying known binding-critical contacts. Second, in a \emph{scoring comparison} experiment, we apply Q-Score to 1000 molecules produced by a 3D molecular generative model across the same 10 targets and compare the resulting rankings against a widely used classical scoring function. This comparison reveals three key findings:
\begin{enumerate}
    \item \textbf{Orthogonality.} Q-Score and classical docking scoring are statistically uncorrelated, demonstrating that Q-Score captures fundamentally different chemical information.
    \item \textbf{Orbital selectivity.} Q-Score is strongly driven by orbital interaction quality and shows no molecular-weight dependence, unlike the classical scoring function.
    \item \textbf{Quantum solver scalability.} DC-QAOA solves the MWVCP across all 1000 instances at 10 qubits with a mean approximation ratio (AR) of 0.94 in simulation (Table~\ref{tab:qaoa_sim}). Hardware execution of 1000 circuits on an IBM Eagle r3 processor shows that 6-qubit instances produce the simulator's most-probable bitstring on 65\% of runs, while 10-qubit instances are noise-limited, establishing a concrete scaling boundary for unmitigated NISQ execution (Table~\ref{tab:hardware}).
\end{enumerate}

\section{Background}
\label{sec:background}

At the molecular scale, biological function depends on physical recognition: a small molecule (the \emph{ligand}) must fit into a specific pocket on a
much larger protein, much as a key fits a lock. The three-dimensional
arrangement of the ligand inside the pocket is called a \emph{binding pose},
and predicting it computationally is the central task of \emph{molecular
docking}. Because candidate drug molecules must be evaluated in the
thousands to millions during early-stage drug design, docking is one of the
most throughput-sensitive workloads in computational biology. At the same
time, pose accuracy depends on faithfully capturing the physics of
molecular interactions, creating a fundamental tension between
computational cost and chemical fidelity.

A docking calculation involves two coupled subproblems: \emph{search} and \emph{scoring}. The search component explores the space of possible ligand
orientations and conformations inside the protein pocket, while the scoring
function evaluates each candidate pose by estimating its interaction
energy. In practice, most widely used docking
tools such as Vina~\cite{https://doi.org/10.1002/jcc.21334} and Glide~\cite{doi:10.1021/jm0306430} employ empirical or statistical
scoring functions that decompose the total energy into a handful of
predefined interaction categories: hydrogen bonds, hydrophobic contacts,
electrostatic terms, and steric penalties~\cite{doi:10.1021/cr00081a009}. This decomposition enables fast evaluation and broad applicability, but it compresses the diversity of real molecular interactions into a small vocabulary. Important electronic
effects, such as charge-transfer donation between lone pairs and
antibonding orbitals, or $\pi$-stacking between aromatic rings, are either conflated with coarser categories or ignored entirely.

One promising direction is to separate docking into a continuous geometric reconstruction problem and a discrete selection problem. In this view, docking can be framed as choosing a small set of interaction anchors that are jointly consistent with a single rigid body placement of the ligand. Each anchor corresponds to a candidate ligand protein contact, while compatibility between anchors is determined by whether they can be satisfied simultaneously under a rigid transformation~\cite{10.1109/99.641604}. This naturally leads to a weighted graph formulation in which anchors form vertices, compatibility defines edges, and the optimal docking hypothesis corresponds to a subset of vertices that is both mutually compatible and high scoring. Identifying such a subset can be expressed as a maximum weight clique objective, which captures the requirement that all selected anchors agree on a single geometric placement.

The Quantum Approximate Optimization Algorithm
(QAOA)~\cite{farhi2014quantumapproximateoptimizationalgorithm} is a hybrid classical-quantum approach
to combinatorial optimization. Given an objective function encoded as a
cost Hamiltonian $H_C$ over $N$ qubits, QAOA prepares a
parameterized quantum state by alternating two operations: a cost unitary $e^{-i\gamma H_C}$ that encodes the problem structure, and a mixer unitary $e^{-i\beta H_M}$ that enables exploration of the solution space. The variational parameters $\{\gamma, \beta\}$ are
optimized by a classical outer loop that minimizes the expectation value
$\langle H_C \rangle$. After optimization, the quantum state is measured
repeatedly, and the most frequently observed bitstring is decoded as the
candidate solution.

A practical limitation of standard QAOA on near-term hardware is that achieving high-quality solutions often requires many alternating layers
(large circuit depth $p$), which accumulates gate errors on noisy
processors~\cite{10.3389/fphy.2014.00005, Harrigan2021}. \emph{Digitized-Counterdiabatic QAOA}
(DC-QAOA)~\cite{PhysRevResearch.4.013141} addresses this by augmenting
each layer with an additional counterdiabatic drive implemented as extra
single-qubit rotations that suppresses transitions away from the target
state. This modification allows DC-QAOA to reach comparable or better
solution quality at lower circuit depths, a critical advantage on current
Noisy Intermediate-Scale Quantum (NISQ) devices where circuit depth is
the primary bottleneck.~\cite{Preskill2018quantumcomputingin} 

\section{Related Work}

\textbf{Quantum approaches to molecular docking.}
The formulation of molecular docking as a maximum-weight vertex clique problem on a Binding Interaction Graph was introduced in~\cite{PhysRevApplied.21.034036}, where QAOA was used to solve the resulting combinatorial optimization. That work relies on manual pharmacophore annotations and a four-category lookup table for graph weights, limiting both automation and chemical resolution. An alternative photonic approach using Gaussian Boson Sampling was proposed in~\cite{doi:10.1126/sciadv.aax1950}. On the data side, QDockBank~\cite{10.1145/3712285.3759799} provides a benchmark of protein fragments predicted on utility-level quantum processors, demonstrating that quantum-generated structures can outperform AlphaFold in docking affinity. Other quantum formulations for biomolecular problems include protein folding~\cite{Robert2021} and drug-target interaction prediction~\cite{https://doi.org/10.1002/advs.202513641}.
Q-Score differs by grounding graph weights in orbital-level quantum chemistry rather than empirical potentials.

\textbf{Quantum-chemistry-informed scoring.}
Classical scoring functions such as Vina~\cite{https://doi.org/10.1002/jcc.21334} and Glide~\cite{doi:10.1021/jm0306430} decompose binding energy into empirical pairwise terms. More physics-based approaches incorporate quantum-mechanical calculations: QM/MM rescoring~\cite{doi:10.1021/acs.chemrev.5b00630},
the fragment molecular orbital method~\cite{C2CP23784A},
and deep learning based method~\cite{McNutt2025}.
These methods improve chemical fidelity but are computationally expensive and not naturally formulated as discrete optimization problems. The SIMG* model~\cite{Boiko2025} used in our pipeline bridges this gap by predicting NBO orbital interactions via a GNN at near-zero marginal cost per complex.

\textbf{QAOA for combinatorial optimization.}
QAOA~\cite{farhi2014quantumapproximateoptimizationalgorithm} and its variants have been benchmarked on problems including MaxCut~\cite{Harrigan2021}, 
portfolio optimization~\cite{Brandhofer2023}, %\cite{TODO_portfolio_qaoa}
and scheduling~\cite{Amaro2022}. %\cite{TODO_scheduling_qaoa}
The digitized-counterdiabatic variant~\cite{PhysRevResearch.4.013141} achieves better approximation ratios at lower circuit depth by augmenting each layer with counterdiabatic driving terms. Hardware demonstrations on superconducting processors have established practical scaling boundaries for unmitigated execution~\cite{Kim2023}.

\section{Method}
\label{sec:method}

\begin{figure*}[t]
    \centering
    \includegraphics[width=\linewidth]{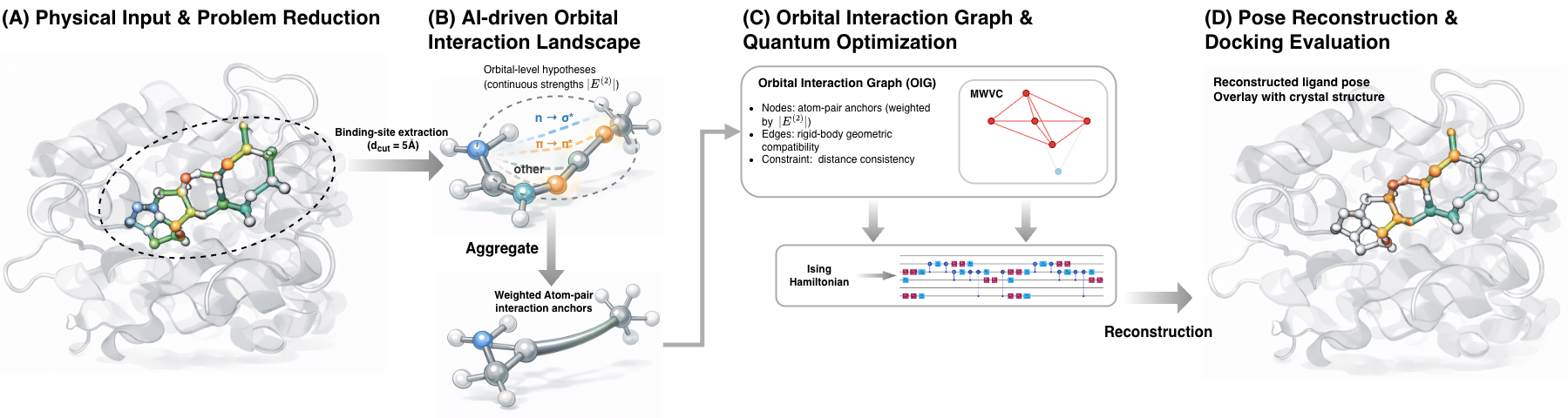}
    \caption{Overview of the hybrid AI–quantum molecular docking pipeline. (A) Starting from a protein–ligand co-crystal structure, the binding site is extracted using a distance-based cutoff to reduce the system size while preserving key interaction context. (B) An AI model predicts orbital-level donor–acceptor interaction hypotheses across the binding site, yielding continuous interaction strengths. These orbital interactions are aggregated into weighted atom-pair interaction anchors. (C) The resulting anchors define an Orbital Interaction Graph (OIG), where nodes represent weighted atom-pair interactions and edges enforce rigid-body geometric compatibility. Docking is formulated as a maximum-weight vertex clique problem and solved using digitized-counterdiabatic QAOA. (D) The selected anchor set is used to reconstruct the ligand pose via rigid-body alignment, and docking accuracy is evaluated by comparison with the experimental crystal structure.}
    \label{fig:overview}
\end{figure*}

This section describes the Q-Score framework, illustrated in Figure~\ref{fig:overview}.
Q-Score proceeds through a sequence of representation transformations: from a protein-ligand 3D structure to orbital-level interaction hypotheses, then to a discrete combinatorial optimization on a weighted graph, and finally to a numerical score and a reconstructed pose if the user requires.

In the \emph{AI phase}, we extract binding interaction data to construct an Orbital Interaction Graph (OIG). We isolate the binding site from a protein-ligand complex using a distance cutoff, employ a Graph Neural Network (GNN) to predict orbital-level donor-acceptor interactions across the complex, and aggregate these into weighted atom-pair anchors that form the OIG nodes.

In the \emph{Quantum phase}, we solve the maximum-weight vertex clique problem (MWVCP) on the OIG via Digitized-Counterdiabatic QAOA (DC-QAOA)~\cite{PhysRevResearch.4.013141}, following the graph-based docking formulation of~\cite{PhysRevApplied.21.034036}. The weight of the optimal clique defines the Q-Score. For redocking validation, we additionally reconstruct the molecular pose using the Kabsch algorithm~\cite{Lawrence_2019} and compute the RMSD against the experimental structure.
\subsection{Binding Site Extraction}
The initial stage of our pipeline involves extracting the binding site atoms from the protein-ligand complex. This extraction is necessitated by the hardware constraints of current Noisy Intermediate-Scale Quantum (NISQ) devices, which lack the capacity to process an entire macromolecular system. Furthermore, since primary binding interactions are localized at the protein-ligand interface, focusing on this region preserves the most relevant chemical information.

Formally, we define the binding site $\mathcal{B}$ as the set of all protein atoms located within a distance cutoff $d_{\mathrm{cut}}$ of any ligand atom:
\begin{equation}
\mathcal{B} = \left\{ a \in \text{protein} \;\middle|\; \min_{l \in \text{ligand}} \| \mathbf{r}_a - \mathbf{r}_l \| < d_{\mathrm{cut}} \right\}
\end{equation}
where $\mathbf{r}_a$ and $\mathbf{r}_l$ denote the 3D coordinates of protein atom $a$ and ligand atom $l$, respectively. We use $d_{\mathrm{cut}} = 5.0$~\AA\ throughout. Across our 11 benchmark targets, this yields reduced complexes of 100-250 atoms that are computationally tractable while capturing the essential binding environment.

\subsection{Orbital Interaction Prediction}
The limitations of previous work~\cite{PhysRevApplied.21.034036} can be summarized as follows. First, the methodology relies on manual pharmacophore identification, where a domain expert assigns interaction types to specific atoms or functional groups. This process is inherently subjective, labor-intensive, and difficult to reproduce across diverse protein-ligand systems. Second, the node weights are derived from a pharmacophore potential lookup table trained on the PDBbind dataset~\cite{doi:10.1126/sciadv.aax1950}, which assigns coarse-grained statistical scores. This discretizes the complex landscape of molecular interactions into only four categories, collapsing the diversity of electronic effects into a narrow, predefined vocabulary. Consequently, critical interactions such as $n \to \sigma^*$ donation, $\pi \to \pi^*$ stacking, and hyperconjugation, which are often essential for binding specificity, have no direct pharmacophore equivalent and are either ignored or conflated with generic categories.

To address these limitations, we replace the manual process with automated, quantum-chemistry-informed orbital interaction prediction via a graph neural network model, SIMG* (Stereoelectronics-Infused Molecular Graphs)~\cite{Boiko2025}. 

Given only the 3D coordinates and connectivity of the ligand-protein complex, SIMG* predicts orbital-level interactions and their second-order perturbation energies $E^{(2)}$ based on Natural Bond Orbital (NBO) analysis~\cite{B1RP90011K, 10.1063/1.449928}. This provides a continuous, physics-grounded energy landscape without human intervention. Because SIMG* is trained on the QM9~\cite{Ramakrishnan2014} and GEOM~\cite{Axelrod2022} datasets, it generalizes effectively across chemical space without requiring system-specific re-fitting of interaction potentials. Furthermore, by operating at the orbital level, explicitly distinguishing lone pairs, bonding orbitals, and antibonding orbitals. The model captures a significantly richer interaction vocabulary than the traditional four-category pharmacophore scheme.

SIMG* operates in two stages. The first stage (\texttt{lp\_pred\_model}) predicts lone pair (LP) counts for each atom from the molecular graph. These predicted lone pairs, together with bond orbitals derived from the connectivity, augment the atomic graph into an orbital-level graph where nodes represent atoms, lone pairs, and bond orbitals. The second stage (\texttt{nbo\_pred\_model}) predicts pairwise orbital interactions and their associated second-order perturbation energies $E^{(2)}$, which quantify the stabilization energy from donor--acceptor orbital overlap:
\begin{equation}
E^{(2)}_{i \to j} = -n_i \frac{|\langle \phi_j | \hat{F} | \phi_i \rangle|^2}{\epsilon_j - \epsilon_i}
\end{equation}
where $\phi_i$ is the donor orbital, $\phi_j$ is the acceptor orbital, $n_i$ is the donor occupancy, $\hat{F}$ is the Fock operator, and $\epsilon_i$, $\epsilon_j$ are the orbital energies.
\subsection{Interaction Graph Construction}

Following the prediction of the full interaction set, we proceed to the construction of the interaction graph. First, we isolate \emph{intermolecular links}, defined as interactions occurring between a ligand orbital and a protein orbital. An orbital $o$ is assigned to either the ligand or protein based on its parent atom index: $o$ belongs to the ligand if $\mathrm{parent}(o) < n_L$. Each intermolecular link is characterized by:
\begin{itemize}
    \item The parent atom indices for the ligand ($a_L$) and protein ($a_P$);
    \item The orbital types (e.g., lone pair (LP), bond (BND), or atomic orbital (ATOM)) for each partner;
    \item The predicted $|E^{(2)}|$ perturbation energy, which serves as the interaction weight.
\end{itemize}
At this stage, all intermolecular links are extracted without an energy threshold to preserve the comprehensive interaction landscape for subsequent processing.

Because a single atom pair may participate in multiple orbital-level interactions—such as a nitrogen lone pair interacting with multiple C--H bond orbitals—we aggregate these links into atom-pair-level interactions by summing their respective orbital energies:
\begin{equation}
    W(a_L, a_P) = \sum_{(o_L, o_P) \in \mathcal{I}(a_L, a_P)} |E^{(2)}_{o_L \to o_P}|
\end{equation}
where $\mathcal{I}(a_L, a_P)$ represents the set of all orbital interaction pairs between ligand atom $a_L$ and protein atom $a_P$. This aggregated energy $W$ defines the potential significance of the contact.

To ensure a tractable qubit count for QAOA while maintaining maximal spatial coverage of the binding site, we employ a greedy diverse selection strategy for node selection. The aggregated atom pairs are sorted by $W(a_L, a_P)$ in descending order and selected subject to two primary constraints:
\begin{enumerate}
    \item \textbf{Uniqueness}: Each ligand or protein atom may appear in only one selected pair, ensuring that each node represents a distinct spatial anchor point.
    \item \textbf{Distance Filtering}: The Euclidean distance $\|\mathbf{r}_{a_L} - \mathbf{r}_{a_P}\|$ must not exceed a maximum cutoff $d_{\max}$, filtering out long-range interactions. We set $d_{\max} = 5.0$~\AA.
\end{enumerate}
This process yields $N$ interaction nodes, where $N$ is a tunable parameter corresponding to the number of qubits. In our evaluation, we test $N \in \{6, 10, 12\}$.

The selected pairs define the nodes of the \emph{Orbital Interaction Graph} (OIG), an undirected weighted graph $G = (V, E, w)$ where:
\begin{itemize}
    \item $V = \{v_1, \ldots, v_N\}$, with each node $v_k$ representing an interaction pair $(a_L^{(k)}, a_P^{(k)})$;
    \item $w_k = W(a_L^{(k)}, a_P^{(k)})$ is the node weight;
    \item $E$ is the set of edges connecting \emph{compatible} node pairs.
\end{itemize}
Two nodes $v_i$ and $v_j$ are connected if they are geometrically compatible under a rigid-body transformation. This is determined by distance consistency:
\begin{equation}
    \left| \, \|\mathbf{r}_{a_L^{(i)}} - \mathbf{r}_{a_L^{(j)}}\| - \|\mathbf{r}_{a_P^{(i)}} - \mathbf{r}_{a_P^{(j)}}\| \, \right| \leq \tau
    \label{eq:compatibility}
\end{equation}
where $\tau$ is the distance tolerance. This criterion ensures that if both interactions are satisfied by a single rigid-body placement, the internal distances between ligand atoms must match those of their protein partners. Nodes sharing identical atom indices are inherently considered compatible.

\begin{figure*}[htbp]
    \centering
    \includegraphics[width=\textwidth]{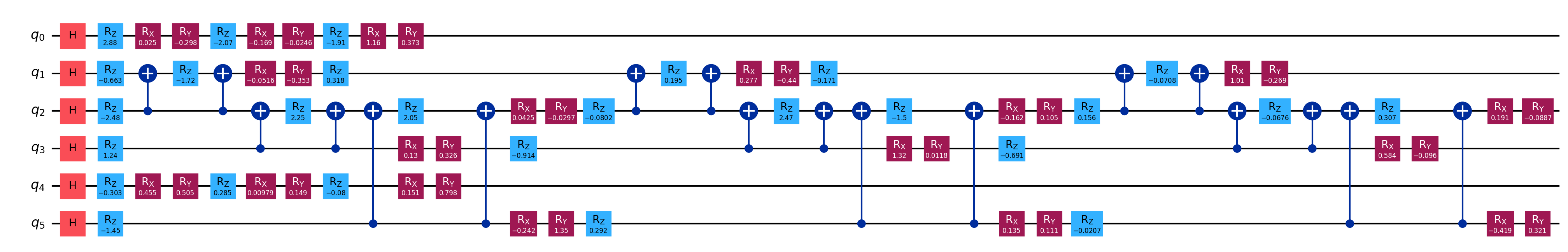}
    \caption{An example DC-QAOA circuit schematic for the 6-qubit, 3-layer configuration.}
    \label{fig:circuit}
\end{figure*}

\subsection{Quantum Optimization via DC-QAOA}

We now transition to the quantum optimization phase of the pipeline. Our objective is to map the molecular docking problem onto the \emph{Maximum Weight Vertex Clique Problem} (MWVCP). In the context of the OIG, a \emph{clique} represents a subset of mutually compatible interaction nodes that can be simultaneously realized by a single rigid-body pose. The MWVCP seeks to identify the clique that maximizes the total interaction energy:
\begin{equation}
    \max_{\mathbf{x} \in \{0,1\}^N} \sum_{k=1}^{N} w_k x_k \quad \text{s.t.} \quad x_i + x_j \leq 1 \;\; \forall (i,j) \notin E
\end{equation}
where $x_k = 1$ indicates the inclusion of interaction node $v_k$ in the clique and correspondingly, $x_k = 0$ indicates the exclusion.

To solve this via quantum annealing or gate-based algorithms, the constrained optimization is converted into a Quadratic Unconstrained Binary Optimization (QUBO) form by introducing a penalty $P$ for constraint violations:
\begin{equation}
    \min_{\mathbf{x}} \left( -\sum_{k=1}^{N} w_k x_k + P \sum_{(i,j) \notin E} x_i x_j \right)
\end{equation}
The penalty $P$ must be sufficiently large to enforce the clique constraint. In our experiments, we set $P = 6.0$ as the default. This value is sufficient when the maximum clique weight is below $P$, which holds for 8 of our 11 redocking targets. Section~\ref{sec:penalty} analyzes the sensitivity of QAOA performance to this parameter.

For execution on a quantum processor, the QUBO is mapped to an Ising Hamiltonian through the substitution $x_k = (1 - Z_k)/2$, where $Z_k$ represents the Pauli-Z operator acting on the $k$-th qubit:
\begin{equation}
    H_C = \sum_{k=1}^{N} h_k Z_k + \sum_{i < j} J_{ij} Z_i Z_j + C
\end{equation}
where the longitudinal fields $h_k$, coupling strengths $J_{ij}$, and the constant offset $C$ are derived directly from the QUBO coefficients.

We solve the MWVCP using DC-QAOA. The DC-QAOA circuit begins with a Hadamard layer to prepare a uniform superposition $|+\rangle^{\otimes N}$, followed by $p$ layers consisting of three distinct unitary operations:

\begin{enumerate}
    \item \textbf{Cost Unitary}: For each Pauli term $P_k$ with coefficient $c_k$ in the Hamiltonian decomposition $H_C = \sum_k c_k P_k$, we apply $\exp(-i \gamma_k c_k P_k)$ using an independent variational parameter $\gamma_k$.
    \item \textbf{Mixer Unitary}: An $R_X(\beta_j)$ rotation is applied independently to each qubit $j \in \{1, \ldots, N\}$.
    \item \textbf{Counterdiabatic Unitary}: An $R_Y(\alpha_j)$ rotation is applied independently to each qubit $j \in \{1, \ldots, N\}$ to mitigate non-adiabatic transitions.
\end{enumerate}

Each Pauli exponential $\exp(-i\theta P_k)$ is decomposed into standard gate sequences comprising basis rotations, CNOT ladders, and $R_Z$ gates. The total number of variational parameters per layer is $|H_C| + 2N$, resulting in $(|H_C| + 2N) \times p$ total parameters across $p$ layers. Notably, our parameterization assigns an independent angle to each Pauli term within the cost layer—a modification from the original DC-QAOA formulation~\cite{PhysRevResearch.4.013141} that typically utilizes a single $\gamma$ per layer.
\subsection{Pose Reconstruction}
Given the optimal clique $\mathcal{C} = \{v_{k_1}, \ldots, v_{k_m}\}$ identified by the quantum optimizer, we reconstruct the 3D ligand pose using the Kabsch algorithm~\cite{Lawrence_2019}. Each node in the clique defines a spatial anchor point, where the ligand atom $a_L^{(k)}$ is mapped to the protein atom $a_P^{(k)}$. Let $\mathbf{P} = [\mathbf{r}_{a_L^{(k_1)}}, \ldots, \mathbf{r}_{a_L^{(k_m)}}]^T$ represent the source coordinates (ligand anchors) and $\mathbf{Q} = [\mathbf{r}_{a_P^{(k_1)}}, \ldots, \mathbf{r}_{a_P^{(k_m)}}]^T$ represent the target coordinates (protein anchors). 

The Kabsch algorithm calculates the optimal rotation matrix $\mathbf{R} \in SO(3)$ and translation vector $\mathbf{t} \in \mathbb{R}^3$ that minimize the root-mean-square deviation (RMSD) between the two sets of points:
\begin{equation}
    \mathrm{RMSD}^2 = \frac{1}{m} \sum_{i=1}^{m} \| \mathbf{R} \mathbf{p}_i + \mathbf{t} - \mathbf{q}_i \|^2
\end{equation}
where $\mathbf{p}_i$ and $\mathbf{q}_i$ denote the centered source and target vectors, respectively. 

The resulting optimal transformation is applied to the coordinates of all atoms in the ligand to generate the predicted 3D pose. Notably, this rigid-body alignment stage addresses a critical gap in previous methodologies~\cite{PhysRevApplied.21.034036}, which typically terminate at the MWVCP solution without generating a physical conformation. By integrating pose reconstruction, we enable the direct evaluation of docking accuracy via RMSD against the experimental crystal structure.

\subsection{Q-Score Definition and Evaluation Metrics}
Given the optimal clique $\mathcal{C}^*$ returned by DC-QAOA, we define the \emph{Q-Score} of a protein-ligand complex as the total clique weight:
\begin{equation}
    \text{Q-Score} = \sum_{v_k \in \mathcal{C}^*} w_k = \sum_{v_k \in \mathcal{C}^*} W(a_L^{(k)}, a_P^{(k)})
\end{equation}
A higher Q-Score indicates a larger set of mutually compatible, high-energy orbital interactions, reflecting stronger stereoelectronic complementarity between the ligand and protein. Unlike classical scoring functions that sum over all atom pairs, Q-Score sums only over the interactions that are jointly realizable under a single rigid-body placement, making it a measure of interaction quality rather than interaction quantity.

Because Q-Score relies on DC-QAOA to solve the underlying MWVCP, the quality of the quantum solver directly affects the scoring output. We evaluate this using the \emph{approximation ratio} (AR), defined as the ratio of the DC-QAOA clique weight to the classical exact optimum. An AR of 1.0 means DC-QAOA found the globally optimal clique; lower values indicate the quantum solver converged to a suboptimal solution. AR thus measures how faithfully the quantum optimizer recovers the intended Q-Score.

To compare how effectively different scoring functions select molecules with strong orbital interactions, we introduce the \emph{Interaction Quality Enrichment Factor} (IQEF). Given a selection method $S$ that picks the top-$K$ molecules from a set, IQEF measures whether $S$ enriches for high orbital-energy molecules relative to random chance:
\begin{equation}
    \text{IQEF}_S = \frac{|\{m \in S : \bar{E}^{(2)}_m > \text{median}(\bar{E}^{(2)})\}| / K}{0.5}
\end{equation}
where $\bar{E}^{(2)}_m$ is the mean $E^{(2)}$ per anchor for molecule $m$. An IQEF of 1.0 means the scorer selects high-$E^{(2)}$ molecules at the same rate as random selection; IQEF $>$ 1.0 indicates the scorer preferentially identifies molecules with strong orbital interactions. By computing IQEF for both Q-Score and a classical scorer on the same molecule set, we can directly test whether each method captures orbital interaction quality.

\section{Experiment Results}
We evaluate Q-Score in two experimental settings: (1)~\emph{redocking} on 11 co-crystal structures, validating that DC-QAOA recovers known binding contacts; and (2)~\emph{scoring comparison} on 1000 AI-generated molecules, demonstrating that Q-Score captures chemical information orthogonal to classical docking. All experiments are run on both a statevector simulator and a 127-qubit IBM Eagle r3 processor (\texttt{ibm\_rensselaer})~\cite{Kim2023}.
\subsection{Experimental Setup}
We test six qubit/layer configurations: $N \in \{6, 10, 12\}$ qubits and $p \in \{3, 6\}$ QAOA layers, corresponding to 60-300 variational parameters. An example circuit is shown in Figure~\ref{fig:circuit}. Throughout, we use a maximum pair distance $d_{\max} = 5.0$~\AA, the COBYLA optimizer~\cite{Powell1994} with 5000 iterations, and a constraint penalty $P = 6.0$. The distance tolerance $\tau$ is set per target (Table~\ref{tab:redocking}).

\textbf{Redocking benchmark.} We apply Q-Score to 11 protein-ligand co-crystal structures: 10 drug targets chosen from a benchmark~\cite{zheng2026affinitybenchmark1d2d} spanning diverse protein families (kinases, GPCRs, nuclear receptors, epigenetic regulators, and immune modulators) plus SARS-CoV-2 main protease (8SKH). For each target, the native ligand is used and the OIG is constructed from the experimental binding pose.

\textbf{Scoring comparison.} We score 1000 molecules (100 per target) generated by Pocket2Mol~\cite{peng2025pocket2molefficientmolecularsampling}, a 3D structure-based generative model. Each molecule is first docked with Vina to obtain a classical score, then processed through the Q-Score pipeline at $N=10$ qubits and $p=3$ layers. We compare the resulting Q-Score and Vina rankings using Spearman correlation, IQEF, and cross-scorer disagreement analysis.
\subsection{SIMG* Interaction Landscape}

For each target, SIMG* predicts thousands of intermolecular orbital interactions from the 3D structure alone. As an illustrative example, the 8SKH complex yields 2496 atom pairs with non-zero aggregated $E^{(2)}$. Only 53 pairs exceed $5.0$~kcal/mol, and the strongest pair: the ligand carbonyl oxygen to CYS145 S$_\gamma$ ($E^{(2)} = 5.79$~kcal/mol), correctly identifies the known covalent binding mechanism without manual annotation. Across all 11 targets, the OIG construction step (Section~\ref{sec:method}) selects $N$ diverse high-energy anchors that span the binding pocket.

\subsection{Redocking Results}

\begin{figure*}[t]
    \centering
    \includegraphics[width=\textwidth]{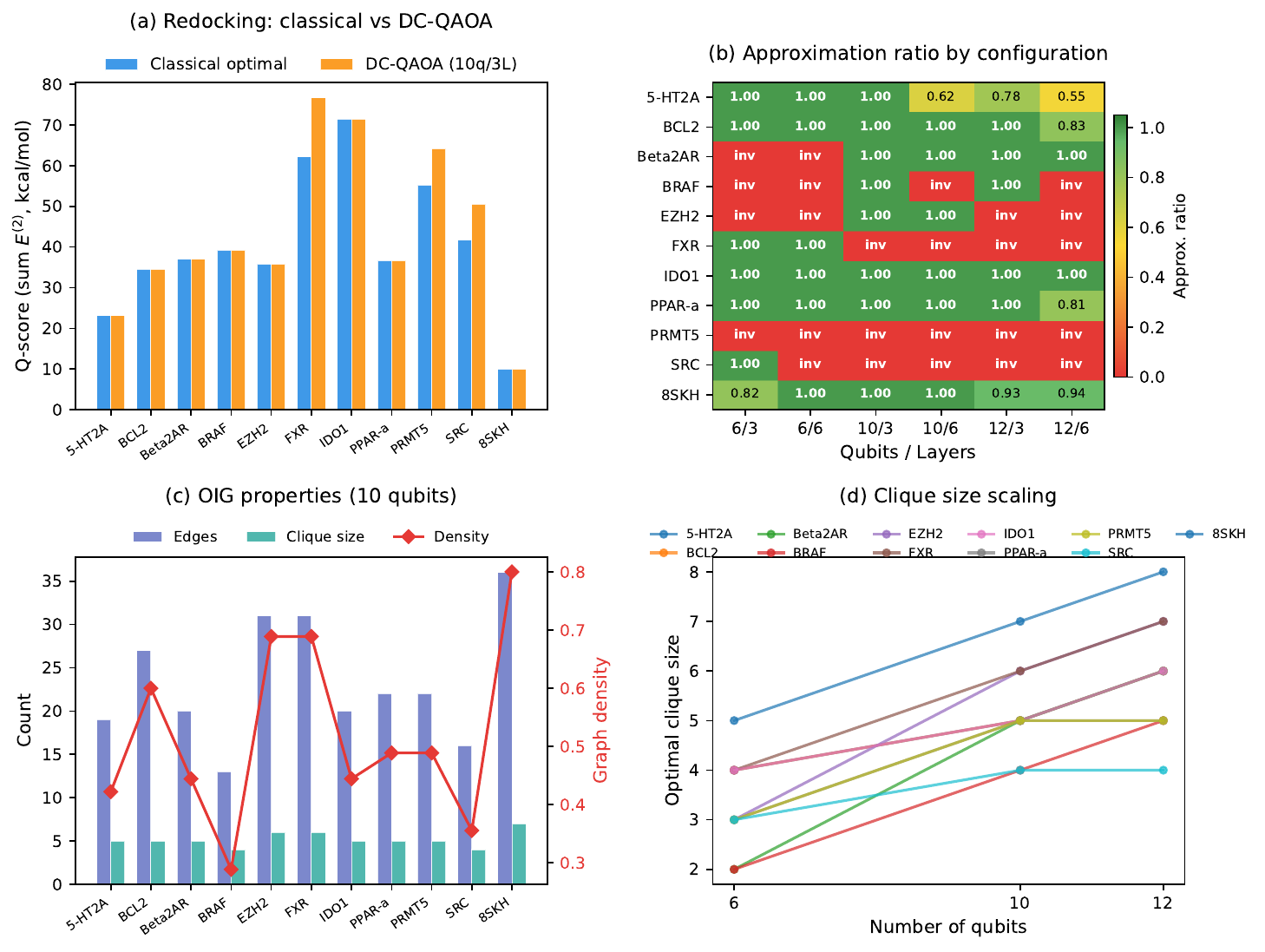}
    \caption{Redocking results across 11 co-crystal structures. (a)~Classical optimal vs DC-QAOA score at 10q/3L. (b)~Approximation ratio heatmap across all configurations; ``inv'' = invalid clique. (c)~OIG structural properties at 10 qubits. (d)~Optimal clique size scaling from 6 to 12 qubits.}
    \label{fig:redocking}
\end{figure*}

\textbf{Multi-target benchmark.} Table~\ref{tab:redocking} and Figure~\ref{fig:redocking} summarize the redocking experiment across all 11 co-crystal structures at the $N=10$, $p=3$ configuration. DC-QAOA recovers the exact classical optimum on 8 out of 11 targets. The three failures (FXR, PRMT5, SRC) return constraint-violating solutions with scores exceeding the classical optimum, indicating that the penalty $P=6$ was insufficient to enforce the clique constraint for these instances. Figure~\ref{fig:redocking}(b) shows the approximation ratio across all six configurations: 10q/3L provides the best trade-off, while 6-qubit problems are often too small to form meaningful cliques, and 12-qubit problems are harder to optimize. The optimal clique size grows from 2--4 at 6 qubits to 4--7 at 12 qubits (Figure~\ref{fig:redocking}(d)), capturing progressively more of the orbital interaction landscape.

\textbf{Penalty sensitivity.}\label{sec:penalty} The three redocking failures have clique weights of 41--71 kcal/mol, exceeding $P=6$ and causing the optimizer to prefer invalid high-weight solutions. A per-target penalty sweep (Table~\ref{tab:penalty}) reveals that $P=10$ recovers the exact optimum on FXR and PRMT5 while SRC reaches AR=0.87. However, $P=10$ degrades targets that succeeded at $P=6$: for example, BRAF drops from AR=1.00 to 0.47. The optimal penalty is therefore \emph{instance-dependent}. Larger penalties enforce feasibility but create a more rugged optimization landscape, a known challenge in penalty-based constrained QAOA.

\begin{table}[t]
\centering
\caption{Penalty sensitivity for the three failed redocking targets (10q/3L). $P{=}6$ yields invalid cliques; $P{=}10$ recovers FXR and PRMT5 but performance degrades for $P{>}15$.}
\label{tab:penalty}
\footnotesize
\begin{tabular}{@{}l rrr@{}}
\toprule
$P$ & FXR (AR) & PRMT5 (AR) & SRC (AR) \\
\midrule
6   & invalid  & invalid    & invalid  \\
10  & \textbf{1.000}  & \textbf{1.000}    & 0.869    \\
15  & 0.807    & 0.413      & 0.378    \\
20  & 0.909    & 0.868      & 0.631    \\
50  & 0.803    & 0.228      & 0.664    \\
\bottomrule
\end{tabular}
\end{table}

% =============================================================
% Table I: Q-score orthogonality analysis
% =============================================================
\begin{table}[t]
\centering
\caption{Q-score captures information orthogonal to classical scoring across 10 drug targets. $\rho$: Spearman rank correlation. IQEF: Interaction Quality Enrichment Factor for the top-25 molecules selected by each scorer.}
\label{tab:qscore_analysis}
\footnotesize
\setlength{\tabcolsep}{3.5pt}
\begin{tabular}{@{}l c r r r r r@{}}
\toprule
Target & $N$
  & $\rho$(V,Q)
  & $\rho$(MW,V)
  & $\rho$($E^{(2)}$\!,Q)
  & IQEF$_\text{V}$
  & IQEF$_\text{Q}$ \\
\midrule
5-HT2A        & 100 & $-$0.04 & $-$0.41 & $+$0.91 & 0.96 & 2.00 \\
BCL2           & 100 & $+$0.01 & $-$0.23 & $+$0.93 & 1.12 & 2.00 \\
Beta2AR        & 100 & $+$0.02 & $-$0.14 & $+$0.96 & 0.80 & 2.00 \\
BRAF           & 100 & $+$0.04 & $-$0.49 & $+$0.89 & 0.80 & 2.00 \\
EZH2           & 100 & $+$0.10 & $-$0.20 & $+$0.79 & 1.28 & 1.84 \\
FXR            & 100 & $+$0.22 & $-$0.03 & $+$0.93 & 0.72 & 2.00 \\
IDO1           & 100 & $+$0.07 & $-$0.05 & $+$0.95 & 0.80 & 2.00 \\
PPAR-$\alpha$  & 100 & $+$0.19 & $-$0.26 & $+$0.89 & 0.72 & 1.92 \\
PRMT5          &  99 & $+$0.08 & $+$0.05 & $+$0.85 & 1.13 & 1.86 \\
SRC            & 100 & $-$0.22 & $-$0.25 & $+$0.91 & 0.96 & 2.00 \\
\midrule
\multicolumn{2}{@{}l}{Mean\,$\pm$\,SEM}
  & $+$0.05\,$\pm$\,0.04
  & $-$0.20\,$\pm$\,0.05
  & $+$0.90\,$\pm$\,0.02
  & 0.93
  & 1.96 \\
\bottomrule
\end{tabular}
\end{table}

% =============================================================
% Table II: QAOA simulation performance
% =============================================================
\begin{table}[t]
\centering
\caption{DC-QAOA simulation performance (10 qubits, 3 layers, COBYLA optimizer) across 1000 molecules from 10 drug targets. AR\,=\,approximation ratio.}
\label{tab:qaoa_sim}
\footnotesize
\setlength{\tabcolsep}{3.5pt}
\begin{tabular}{@{}l c r r r r@{}}
\toprule
Target & $N$ & Mean AR & Exact\,(\%) & AR${\geq}$0.9\,(\%) & AR${\geq}$0.8\,(\%) \\
\midrule
5-HT2A        & 100 & 0.943 & 52.0 & 77.0 & 89.0 \\
BCL2           & 100 & 0.907 & 34.0 & 74.0 & 86.0 \\
Beta2AR        &  98 & 0.938 & 45.9 & 78.6 & 90.8 \\
BRAF           &  97 & 0.908 & 45.4 & 73.2 & 86.6 \\
EZH2           &  99 & 0.967 & 53.5 & 87.9 & 99.0 \\
FXR            & 100 & 0.965 & 67.0 & 89.0 & 93.0 \\
IDO1           & 100 & 0.943 & 53.0 & 84.0 & 93.0 \\
PPAR-$\alpha$  & 100 & 0.966 & 64.0 & 88.0 & 94.0 \\
PRMT5          &  97 & 0.898 & 35.1 & 73.2 & 84.5 \\
SRC            & 100 & 0.980 & 68.0 & 96.0 & 99.0 \\
\midrule
All            & 991 & 0.942 & 51.9 & 82.1 & 91.5 \\
\bottomrule
\end{tabular}
\end{table}

% =============================================================
% Table: Hardware execution summary
% =============================================================
\begin{table}[t]
\centering
\caption{Hardware execution on IBM Eagle r3 (\texttt{ibm\_rensselaer}), 10000 shots per circuit. 250 molecules per configuration (25 per target\,$\times$\,10 targets). Match\,=\,hardware most-probable bitstring equals simulator.}
\label{tab:hardware}
\footnotesize
\setlength{\tabcolsep}{4pt}
\begin{tabular}{@{}l r r r r r@{}}
\toprule
Config & HW AR & Sim AR & Match\,(\%) & Valid\,(\%) & $P_\text{opt}$ \\
\midrule
6q/3L  & 0.744 & 1.002 & 66.0 & 25.6 & 0.522 \\
6q/6L  & 0.704 & 1.005 & 64.8 & 30.0 & 0.468 \\
10q/3L & 0.711 & 0.990 & 14.4 &  9.6 & 0.093 \\
10q/6L & 0.688 & 0.987 & 12.4 & 11.6 & 0.069 \\
\bottomrule
\end{tabular}
\end{table}

% =============================================================
% Table: Per-target hardware results
% =============================================================
\begin{table}[t]
\centering
\caption{Per-target hardware approximation ratio and bitstring match rate on IBM Eagle. Each cell aggregates 50 jobs (25 molecules\,$\times$\,2 layer depths).}
\label{tab:hardware_per_target}
\footnotesize
\setlength{\tabcolsep}{4pt}
\begin{tabular}{@{}l rr rr@{}}
\toprule
 & \multicolumn{2}{c}{Mean HW AR} & \multicolumn{2}{c}{Match\,(\%)} \\
\cmidrule(lr){2-3} \cmidrule(lr){4-5}
Target & 6q & 10q & 6q & 10q \\
\midrule
5-HT2A        & 0.742 & 0.653 & 68.0 & 18.0 \\
BCL2           & 0.674 & 0.643 & 52.0 &  8.0 \\
Beta2AR        & 0.656 & 0.660 & 66.0 &  8.0 \\
BRAF           & 0.653 & 0.772 & 26.0 &  0.0 \\
EZH2           & 0.790 & 0.853 & 50.0 &  0.0 \\
FXR            & 0.701 & 0.713 & 92.0 & 24.0 \\
IDO1           & 0.767 & 0.689 & 82.0 & 20.0 \\
PPAR-$\alpha$  & 0.726 & 0.670 & 82.0 & 12.0 \\
PRMT5          & 0.717 & 0.658 & 48.0 & 10.0 \\
SRC            & 0.811 & 0.681 & 88.0 & 34.0 \\
\midrule
Mean           & 0.724 & 0.699 & 65.4 & 13.4 \\
\bottomrule
\end{tabular}
\end{table}

% =============================================================
% Table: Redocking results
% =============================================================
\begin{table}[t]
\centering
\caption{Redocking results on 11 co-crystal structures (10 qubits, 3 layers). Classical score is the exact MWVCP solution. $\checkmark$\,=\,DC-QAOA found the exact optimum. $\times$\,=\,QAOA returned an invalid clique (constraint violation).}
\label{tab:redocking}
\footnotesize
\setlength{\tabcolsep}{2.8pt}
\begin{tabular}{@{}ll r r r r r c@{}}
\toprule
Target & PDB & $\tau$ & Edges & Density & Classical & QAOA & Match \\
\midrule
5-HT2A        & 7WC7 & 5.5 & 19 & 0.42 & 23.11 & 23.11 & $\checkmark$ \\
BCL2           & 6GL8 & 5.5 & 27 & 0.60 & 34.59 & 34.59 & $\checkmark$ \\
Beta2AR        & 8JJL & 5.0 & 20 & 0.44 & 37.02 & 37.02 & $\checkmark$ \\
BRAF           & 1UWH & 3.5 & 13 & 0.29 & 39.21 & 39.21 & $\checkmark$ \\
EZH2           & 5WFD & 5.0 & 31 & 0.69 & 35.78 & 35.78 & $\checkmark$ \\
FXR            & 7D42 & 6.5 & 31 & 0.69 & 62.18 & 76.67 & $\times$ \\
IDO1           & 6AZV & 5.5 & 20 & 0.44 & 71.47 & 71.47 & $\checkmark$ \\
PPAR-$\alpha$  & 1KKQ & 5.0 & 22 & 0.49 & 36.67 & 36.67 & $\checkmark$ \\
PRMT5          & 7S1S & 4.5 & 22 & 0.49 & 55.27 & 64.20 & $\times$ \\
SRC            & 7OTE & 5.0 & 16 & 0.36 & 41.69 & 50.58 & $\times$ \\
8SKH           & 8SKH & 6.0 & 36 & 0.80 &  9.96 &  9.96 & $\checkmark$ \\
\bottomrule
\end{tabular}
\end{table}

% =============================================================
% Table: Simulation cost
% =============================================================
\begin{table}[t]
\centering
\caption{Simulation computational cost. Wall-clock measured on a single CPU core (COBYLA, 5000 iterations) for the BRAF redocking instance.}
\label{tab:sim_cost}
\small
\begin{tabular}{@{}l r r r@{}}
\toprule
Config & Params & Pauli terms & Wall-clock (s) \\
\midrule
6q/3L  &  78 & 14 &  55 \\
6q/6L  & 156 & 14 & 100 \\
10q/3L & 156 & 32 & 275 \\
10q/6L & 312 & 32 & 533 \\
12q/3L & 240 & 56 & 394 \\
12q/6L & 480 & 56 & 759 \\
\bottomrule
\end{tabular}
\end{table}

% =============================================================
% Table: Hardware circuit cost
% =============================================================
\begin{table}[t]
\centering
\caption{Hardware circuit metrics on \texttt{ibm\_rensselaer}. Depth and gates are after transpilation (mean $\pm$ std over 250 circuits). QPU time is per circuit (10000 shots).}
\label{tab:hw_cost}
\small
\begin{tabular}{@{}l r r r@{}}
\toprule
Config & Transpiled depth & Transpiled gates & QPU time (s) \\
\midrule
6q/3L  & $90\pm71$   & $185\pm124$  & 4.0 \\
6q/6L  & $172\pm146$ & $343\pm270$  & 4.0 \\
10q/3L & $285\pm166$ & $710\pm409$  & 4.0 \\
10q/6L & $566\pm344$ & $1416\pm866$ & 4.3 \\
\bottomrule
\end{tabular}
\end{table}

\textbf{8SKH case study.} We evaluate all six configurations on 8SKH as case study. At 6 qubits, DC-QAOA achieves AR = 0.82 (3L, success probability 50.27\%) and AR = 1.00 (6L, 98.85\%), showing that deeper circuits recover the global optimum for small instances. At 10 qubits, both 3L and 6L achieve AR = 1.00, with success probabilities of 69.25\% and 54.44\% respectively; the lower probability at 6L reflects the higher-dimensional parameter landscape. At 12 qubits (240--480 parameters), the optimizer converges to local minima with AR $\approx$ 0.93. Pose reconstruction via Kabsch alignment yields all-atom RMSD of 3.34--3.45~\AA\ across configurations, with 100\% of the 45 ligand atoms within 5~\AA\ of the crystal structure and up to 31.1\% within 3~\AA\ at 12 qubits.

\subsection{Scoring Comparison: Q-Score vs Classical Docking}
We now evaluate Q-Score as a scoring function by comparing its molecule rankings against Vina across 1000 generated molecules.

\begin{figure*}[t]
    \centering
    \includegraphics[width=\textwidth]{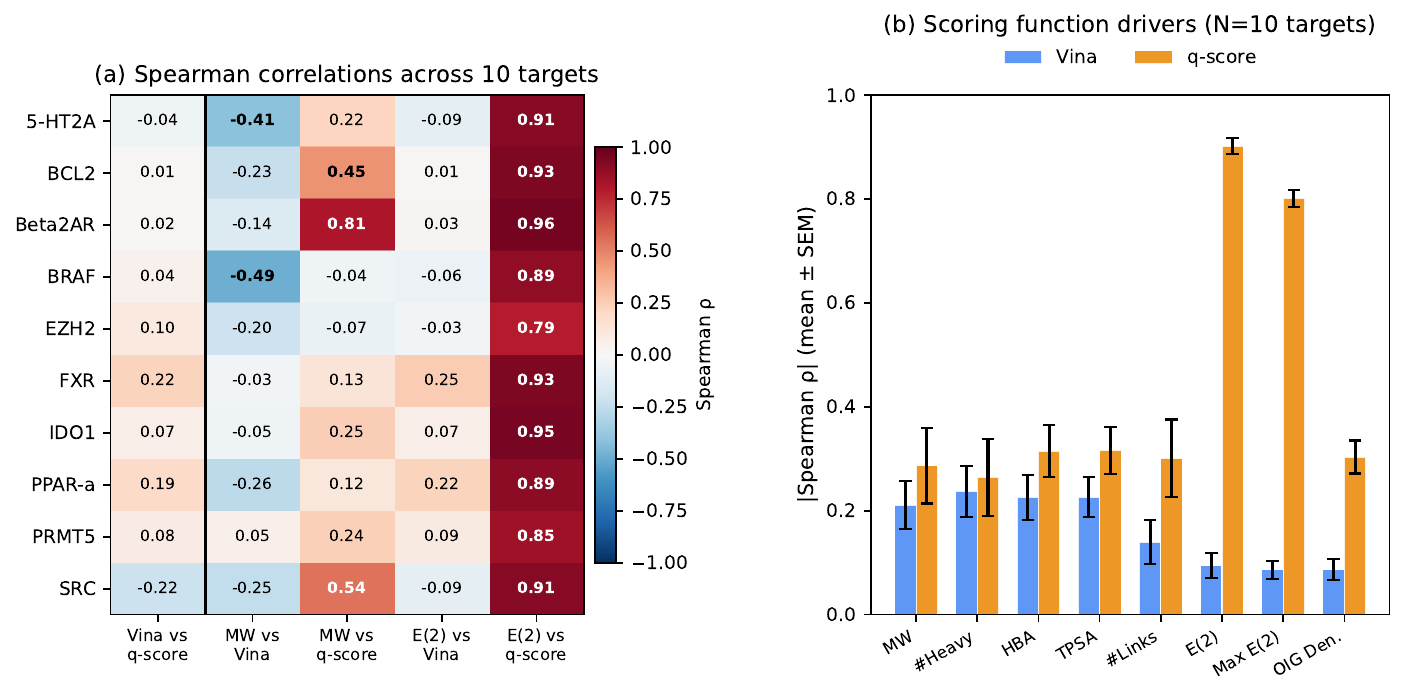}
    \caption{Q-Score captures orthogonal chemical information. (a)~Spearman correlation heatmap across 10 targets: Q-Score is uncorrelated with Vina ($\rho \approx 0$), strongly correlated with orbital energies $E^{(2)}$ ($\rho = 0.90$), and independent of molecular weight. Vina correlates with molecular weight ($\rho = -0.20$). (b)~Mean $|\rho|$ across 10 targets: Vina is driven by size-related properties while Q-Score is driven by $E^{(2)}$.}
    \label{fig:correlation}
\end{figure*}

\textbf{Orthogonality.} Table~\ref{tab:qscore_analysis} and Figure~\ref{fig:correlation}(a) show that Q-Score and Vina are statistically uncorrelated: the mean Spearman $\rho$ across 10 targets is $0.05 \pm 0.04$. The two scorers agree on fewer than 12--28\% of their top-25 selections, indicating that they capture fundamentally different aspects of protein-ligand recognition.

\textbf{Orbital selectivity.} Figure~\ref{fig:correlation}(b) reveals the mechanistic difference. Vina correlates most strongly with molecular weight ($|\rho| = 0.20$), heavy atom count, and polar surface area---all size-related descriptors. Q-Score shows near-zero correlation with these properties but correlates strongly with mean $E^{(2)}$ per anchor ($\rho = 0.90$) and max $E^{(2)}$ ($\rho = 0.80$). This confirms that Q-Score is driven by orbital interaction quality, not molecular bulk.

\begin{figure*}[t]
    \centering
    \includegraphics[width=\textwidth]{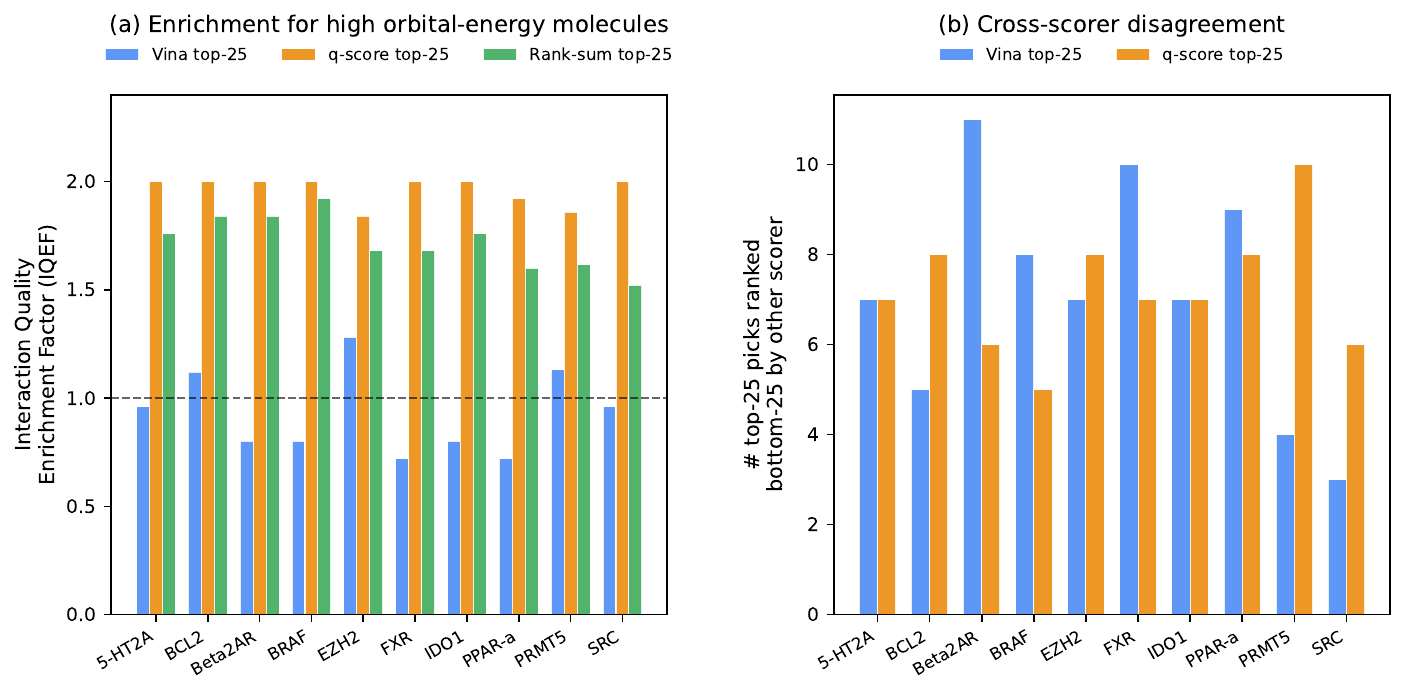}
    \caption{(a)~Interaction Quality Enrichment Factor: Q-Score enriches for high-$E^{(2)}$ molecules at $2\times$ the rate of random selection; Vina provides no enrichment. (b)~Cross-scorer disagreement: each scorer's top-25 contains 4--11 molecules ranked bottom-25 by the other.}
    \label{fig:iqef}
\end{figure*}

\textbf{Enrichment and consistency.} Figure~\ref{fig:iqef}(a) shows the IQEF across all 10 targets. Q-Score achieves a mean IQEF of 1.96, nearly $2\times$ the rate of random selection for high orbital-energy molecules while Vina's IQEF is 0.93, which is indistinguishable from random. We note that the strong Q-Score/$E^{(2)}$ correlation ($\rho = 0.90$) and high IQEF are expected by construction, since Q-Score is defined as a weighted sum of $E^{(2)}$ energies; these metrics confirm internal consistency rather than independent docking validation. The non-trivial finding is the \emph{orthogonality}: this orbital-level signal is statistically uncorrelated with classical scoring ($\rho = 0.05$) and independent of molecular weight ($\rho \approx 0$), demonstrating that Q-Score encodes genuinely different chemical information. Figure~\ref{fig:iqef}(b) quantifies cross-scorer disagreement: 71 out of 250 Vina top-25 picks are ranked bottom-25 by Q-Score, and vice versa.

\textbf{QAOA simulation at scale.} Table~\ref{tab:qaoa_sim} reports DC-QAOA performance across all 1000 generated molecules at $N=10$, $p=3$. The mean approximation ratio is 0.94, with 52\% of instances reaching the exact optimum and 92\% achieving AR $\geq 0.8$. Performance is consistent across all 10 targets, with per-target mean AR ranging from 0.90 (PRMT5) to 0.98 (SRC).

\subsection{Hardware Results}

\begin{figure*}[t]
    \centering
    \includegraphics[width=\textwidth]{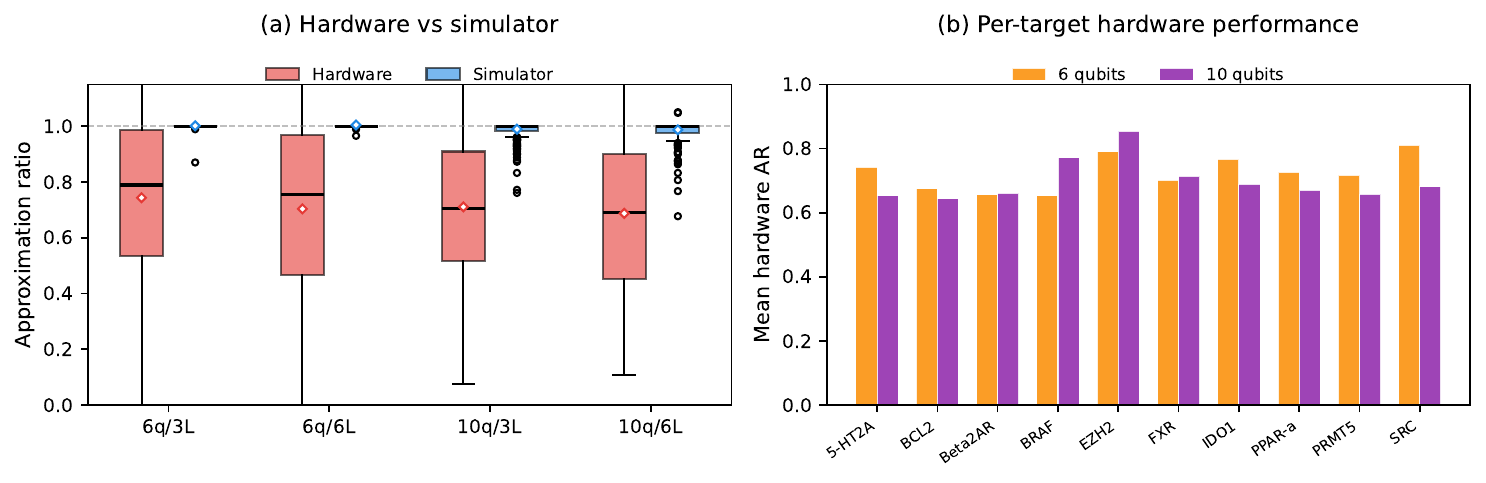}
    \caption{Hardware results on IBM Eagle r3 (1000 circuits, 10000 shots each). (a)~Hardware vs simulator approximation ratio by configuration. (b)~Per-target mean hardware AR for 6 and 10 qubits.}
    \label{fig:hardware}
\end{figure*}

We executed 1000 DC-QAOA circuits on the IBM Eagle r3 processor (\texttt{ibm\_rensselaer})~\cite{Kim2023} using 10000 shots per circuit, covering 25 molecules per target $\times$ 10 targets $\times$ 4 configurations (6q/3L, 6q/6L, 10q/3L, 10q/6L). Table~\ref{tab:hardware} and Figure~\ref{fig:hardware} summarize the results. In the table, Match denotes the fraction of instances where the hardware's most-probable bitstring equals the simulator's, and Valid denotes the fraction where the decoded bitstring forms a valid clique (all selected nodes are pairwise compatible).

\textbf{6-qubit.} At 6 qubits, the hardware matches the simulator's most-probable bitstring on 65\% of instances, with a mean AR of 0.72 and a mean success probability of 0.50. Increasing from 3 to 6 layers slightly reduces performance (match rate 66\% $\to$ 65\%, probability 0.52 $\to$ 0.47), consistent with noise accumulation in deeper circuits.

\textbf{10-qubit.} At 10 qubits, the match rate drops to 13\% and success probability falls to 0.08, indicating that incoherent noise overwhelms the QAOA signal. The hardware AR remains 0.70, close to the 6-qubit value, suggesting that the hardware still concentrates probability on reasonably high-quality bitstrings even when it cannot recover the exact simulator solution.

\textbf{Cross-target consistency.} Figure~\ref{fig:hardware}(b) shows that the 6q $>$ 10q performance gap is consistent across all 10 targets, confirming that the noise scaling is systematic rather than target-dependent. This establishes a concrete boundary: at 6 qubits, the hardware produces the simulator's most-probable bitstring on 65\% of instances with mean AR of 0.72, while at 10 qubits, the match rate drops to 13\%, indicating that error mitigation or improved hardware fidelity is needed to scale beyond 6 qubits on current Eagle-class processors.
\begin{figure*}[htbp]
    \centering
    \includegraphics[width=\textwidth]{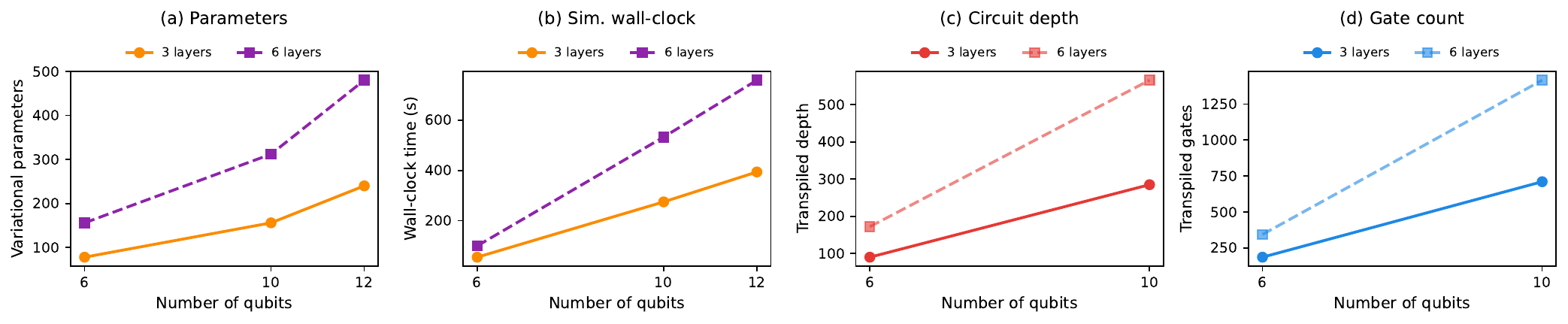}
    \caption{Computational cost and scaling. \textbf{Simulation}: (a)~variational parameter count and (b)~wall-clock time on a single CPU across 6, 10, and 12 qubits. \textbf{Hardware}: (c)~transpiled circuit depth and (d)~gate count on IBM Eagle r3 for 6 and 10 qubits. Solid = 3 layers, dashed = 6 layers.}
    \label{fig:scaling_proj}
\end{figure*}
\subsection{Computational Cost and Scaling}

\textbf{Simulation and hardware cost.} Figure~\ref{fig:scaling_proj} summarizes the computational scaling. The number of Pauli terms grows quadratically with qubit count (14 at 6q, 32 at 10q, 56 at 12q), driving variational parameters from 78 to 480 and wall-clock time from 55s to 759s on a single AMD Threadripper PRO 5955WX CPU core. Across 1000 generated molecules at 10q/3L, the mean pipeline time is 235s. Transpiled circuit depth on \texttt{ibm\_rensselaer} grows from 90 at 6q/3L to 566 at 10q/6L; gate counts scale from 185 to 1416. This depth scaling directly explains the hardware noise gap. QPU time is constant at ${\sim}$4s per circuit.

\textbf{Classical baselines.} At 10 qubits, classical solvers trivially optimize the MWVCP. On 250 benchmark instances, exact enumeration and simulated annealing both achieve 100\% mean AR, greedy selection achieves 99.4\%, and DC-QAOA achieves 96.9\%. The greedy heuristic fails to find the exact optimum on 8.8\% of instances, while DC-QAOA misses on 32.4\% but achieves AR $\geq$ 0.8 on 97.6\%. The MWVCP is NP-hard in general: on real OIG instances, classical exact time grows from 0.1~ms at $N{=}10$ to 102~s at $N{=}250$, while QAOA parameters grow polynomially as $O(N^2 p)$.

\section{Conclusion}
We introduced Q-Score, a quantum-native scoring function for molecular docking that replaces empirical pairwise contact sums with orbital-level donor-acceptor interaction energies solved via combinatorial optimization. By encoding GNN-predicted $E^{(2)}$ energies into an Orbital Interaction Graph and solving the resulting MWVCP with DC-QAOA, Q-Score captures stereoelectronic information that classical scoring functions cannot access.

Across 10 drug targets and 1000 AI-generated molecules, Q-Score is orthogonal to classical docking, driven by orbital interaction quality, and free of molecular-weight bias. The Q-Score/$E^{(2)}$ correlation is expected by construction; the non-trivial finding is that this orbital signal is uncorrelated with classical scoring and molecular size. In redocking on 11 co-crystal structures, DC-QAOA recovers the exact optimum on 8 of 11 targets with $P{=}6.0$, with penalty sensitivity analysis showing the optimal $P$ is instance-dependent. Simulation across 1000 instances yields mean AR of 0.94, with 52\% exact. Classical solvers achieve equivalent quality at 10 qubits, but classical exact time grows exponentially on larger OIG instances (0.1ms at 10 nodes to 102s at 250 nodes), while QAOA resources scale polynomially. Hardware execution of 1000 circuits on IBM Eagle shows 65\% bitstring match at 6 qubits and 13\% at 10 qubits.

Future directions include error mitigation~\cite{PhysRevX.7.021050, PhysRevLett.119.180509} to extend the hardware frontier beyond 6 qubits, adaptive penalty scheduling, and scaling the OIG to 20--50 nodes through finer orbital resolution to test the projected quantum advantage.
\clearpage
\bibliographystyle{IEEEtran}
\bibliography{IEEEabrv,sample-base}

\end{document}